\documentclass[aps,prd,preprint,preprintnumbers,nofootinbib]{revtex4}

\usepackage{graphicx}  
\usepackage{dcolumn}
\usepackage{bm}
\usepackage{amsmath,amssymb,amsfonts}
\usepackage{latexsym}
\usepackage{color}
\usepackage[colorlinks=true,urlcolor=blue,anchorcolor=blue
,citecolor=blue,filecolor=blue,linkcolor=blue,menucolor=blue
,linktocpage=true,pdfproducer=medialab,pdfa=true]{hyperref}
\def\nn    {\nonumber}

\begin{document}
\title{\boldmath
Probing an extra Higgs boson at future linear $e^+ e^-$ colliders}
\author{Wei-Shu Hou}
\email[E-mail address: ]
{wshou@phys.ntu.edu.tw}
\author{Mohamed Krab}
\email[E-mail address: ]
{mkrab@hep1.phys.ntu.edu.tw}
\affiliation{Department of Physics, National Taiwan University, Taipei 10617, Taiwan}
%
\begin{abstract}
We investigate the possibility of probing an extra Higgs boson at future linear $e^+ e^-$ colliders. We consider the production process $e^+e^- \to H\nu \bar\nu$, followed by the decay $H \to W^+W^-$, where $H$ is the extra {\it CP}-even Higgs boson of the general two Higgs doublet model (G2HDM). This process is governed by the {\it CP}-even Higgs mixing angle, $\cos\gamma$, offering direct access to this parameter. 
We discuss constraints on $\cos\gamma$ using existing LHC data and test the viability of the G2HDM top-quark-driven scenario for electroweak baryogenesis. 
We perform a full Monte Carlo simulation of the signal and background, and show that an extra Higgs boson in the mass range $200 \leq m_H \leq 400$~GeV could be probed at high energy linear $e^+ e^-$ colliders.
Promising results are found for CLIC running at 1.5 and 3~TeV collision energies.
\end{abstract}
\maketitle
\section{Introduction}
To date, ATLAS and CMS experiments at the Large Hadron Collider (LHC) have provided stringent tests of the Standard Model (SM) Higgs sector through precision measurements and direct searches for additional scalars.
These results strongly constrain many extensions of the SM, yet simple scenarios that link the Higgs sector to new physics remain viable.
In particular, the Two Higgs Doublet Model (2HDM)
offers a simple framework in which the SM scalar sector is extended by an additional Higgs doublet, leading to new scalar states, besides the observed 125~GeV state by ATLAS~\cite{ATLAS:2012yve} and CMS~\cite{CMS:2012qbp}. 

Interestingly, Ref.~\cite{Fuyuto:2017ewj} has shown that the extra top Yukawa coupling $\rho_{tt}$ of the general 2HDM (G2HDM) can generate baryon asymmetry of the Universe, thereby making it a motivated parameter to probe experimentally. 
Another important parameter worth testing at collider experiments is the $h$-$H$ mixing angle $c_\gamma \equiv \cos\gamma$, with $h$ the observed lightest {\it CP}-even Higgs (usually assumed to be the SM-like Higgs), and $H$ is the extra {\it CP}-even Higgs boson. 
The G2HDM allows for flavor-changing neutral Higgs (FCNH) decays, such as $t \to ch$~\cite{Hou:1991un,Chen:2013qta}.
The absence of any experimental observation of this decay can be attributed to the smallness of $c_\gamma$, as required by SM Higgs properties, which effectively suppresses flavor-changing SM-like Higgs decays without invoking the usual $Z_2$ symmetry. 
Both ATLAS~\cite{ATLAS:2024mih} and CMS~\cite{CMS:2024ubt} have searched for the $t \to ch$ decay and established upper limits on its branching ratio. These limits are interpreted~\cite{Hou:2024bzh,Krab:2025zuy} to constrain the FCNH coupling $\rho_{tc}$ and $c_\gamma$.
In this paper, we first test the model and constrain $\rho_{tt}$ and $c_\gamma$ using LHC data on SM Higgs properties as well as searches for additional Higgs bosons. 
We then study the discovery prospects for the extra Higgs boson $H$ at future linear $e^+ e^-$ colliders. We consider the $e^+e^- \to H\nu \bar\nu$ production process, followed by the $H \to W^+ W^-$ decay. This process is controlled by $c_\gamma$, thereby offering a direct probe of this parameter. We show that a future high energy linear $e^+ e^-$ collider, such as the Compact Linear Collider (CLIC)~\cite{Linssen:2012hp,Aicheler:2018arh}, would be able to probe $m_H$ in the range 200-400~GeV.

Compared to LHC, the future $e^+e^-$ colliders are a cleaner collision environment. An environment free of QCD background, providing a rich program of Higgs physics.
Multiple $e^+e^-$ Higgs factories have been proposed, including CLIC, the International Linear Collider (ILC)~\cite{ILCInternationalDevelopmentTeam:2022izu}, the Future Circular Collider (FCC)~\cite{FCC:2025lpp}, and the Circular Electron Positron Collider (CEPC)~\cite{CEPCPhysicsStudyGroup:2022uwl}. 
Among these colliders, we choose CLIC with its 1.5 and 3 TeV stages. 
With these large center of mass energies, besides increasing precision in Higgs and top quark physics~\cite{Abramowicz:2016zbo,CLICdp:2018esa}, CLIC is also able to directly search for new physics beyond the SM~\cite{CLIC:2018fvx}. Here, we investigate the possibility of probing a heavy Higgs boson at CLIC. 
See, e.g., Refs.~\cite{ Cheung:2023qnj,Hashemi:2023tej,Lee:2025hgb}, for studies about searching for extra Higgs bosons at CLIC.

This paper is organized as follows. We present the theoretical framework in section~\ref{sect:model}. We discuss the LHC constraints in section~\ref{sect:constraints}. We perform a collider analysis in section~\ref{sect:collider}, and we conclude in section~\ref{sect:concl}.

\section{Theoretical Framework}\label{sect:model}
We consider the G2HDM that introduces two identical hypercharge Higgs doublets. The most general potential of a {\it CP}-conserving model can be written in the Higgs basis as~\cite{Davidson:2005cw,Hou:2017hiw}
\begin{align}
V(\Phi,\Phi') =& ~\mu_{11}^2|\Phi|^2 + \mu_{22}^2|\Phi'|^2
    - (\mu_{12}^2\Phi^\dagger\Phi' + \rm{H.c.}) \nn\\
 & + \frac{\eta_1}{2}|\Phi|^4 + \frac{\eta_2}{2}|\Phi'|^4
   + \eta_3|\Phi|^2|\Phi'|^2  + \eta_4 |\Phi^\dagger\Phi'|^2 \nn \\
 & + \left[\frac{\eta_5}{2}(\Phi^\dagger\Phi')^2
   + \left(\eta_6 |\Phi|^2 + \eta_7|\Phi'|^2\right) \Phi^\dagger\Phi' + \rm{H.c.}\right],
\label{pot}
\end{align}
where the Higgs doublet $\Phi$ acquires a vacuum expectation value $\mathit{v}$ through electroweak symmetry breaking (EWSB), while $\left\langle \Phi'\right\rangle = 0$ (hence $\mu_{22}^2 > 0$).
The parameters $\eta_{1\text{--}7}$ in Eq.~\ref{pot} are the Higgs quartic couplings\footnote{It should be noted that $\eta_6$ and $\eta_7$ would be absent in the 2HDM with softly broken $Z_2$ symmetry.} and are real. After EWSB, the Higgs masses can be written in terms of the potential parameters in Eq.~(\ref{pot}),
\begin{align}
  m_{h,H}^2 &= \frac{1}{2}\bigg[m_A^2 + (\eta_1 + \eta_5) v^2 \mp \sqrt{\left(m_A^2+ (\eta_5 - \eta_1) v^2\right)^2
   + 4 \eta_6^2 v^4}\bigg], \\
  m_{A}^2 &= \mu_{22}^2 +  \frac{1}{2}(\eta_3 + \eta_4 - \eta_5) v^2,\\	
  m_{H^\pm}^2 &= \mu_{22}^2 +  \frac{1}{2}\eta_3 v^2.  
\end{align}

The Higgs-fermion couplings are given by \cite{Davidson:2005cw}
\begin{align}
\mathcal{L}_Y = 
& - \frac{1}{\sqrt{2}} \sum_{f = u, d, \ell}
\bar f_{i} \bigg[\big(\lambda^f_{ij} s_\gamma + \rho^f_{ij} c_\gamma\big)h 
 + \big(\lambda^f_{ij} c_\gamma - \rho^f_{ij} s_\gamma\big) H
-i\,{\rm sgn}(Q_f) \rho^f_{ij} A \bigg]  P_R f_{j}\nn \\
& - \bar{u}_i\big[(V\rho^d)_{ij} P_R - (\rho^{u\dagger}V)_{ij} P_L \big]d_j H^+ 
 - \bar{\nu}_i\rho^\ell_{ij} P_R \ell_j H^+ +{\rm H.c.},
\label{LYukawa}
\end{align}
where $i,j = 1\text{--}3$ are the generation indices, $P_{L,R} = (1\mp\gamma_5)/2$, $c_\gamma$ is the $h\text{--}H$ mixing angle (with $\gamma$ corresponding to $\beta-\alpha$ in Ref.~\cite{Davidson:2005cw}), $s_\gamma \equiv \sin\gamma$, 
and $V$ denotes the CKM matrix. 
The matrices $\lambda^f_{ij}$ are real and are given by $\lambda^f_{ij} \equiv \delta_{ij}\sqrt{2}m_i^f/v$ (diagonal), while the matrices $\rho^f_{ij}$ are, in general,
complex and non-diagonal. Here, $\rho^f_{ij}$ are taken to be real, since we assume a {\it CP}-invariant G2HDM. In what follows, we drop the superscript $f$.
We consider the G2HDM top-quark-driven scenario for electroweak baryogenesis (EWBG)~\cite{Fuyuto:2017ewj}, and constrain $\rho_{tt}$ and $c_\gamma$ using LHC data, including measurements of SM Higgs properties and searches for additional Higgs bosons.

\section{LHC constraints}
\label{sect:constraints}
Before turning to the experimental bounds from LHC data, it is useful to recall the theoretical requirements that restrict the parameter space of the G2HDM. These include perturbativity of the quartic couplings, tree-level unitarity of the scalar-scalar scattering amplitudes, and vacuum stability to ensure a bounded-from-below scalar potential. These conditions limit the quartic couplings in Eq. \ref{pot} and Higgs masses. 

\begin{figure}[t!]
	\centering
	\includegraphics[scale=0.55]{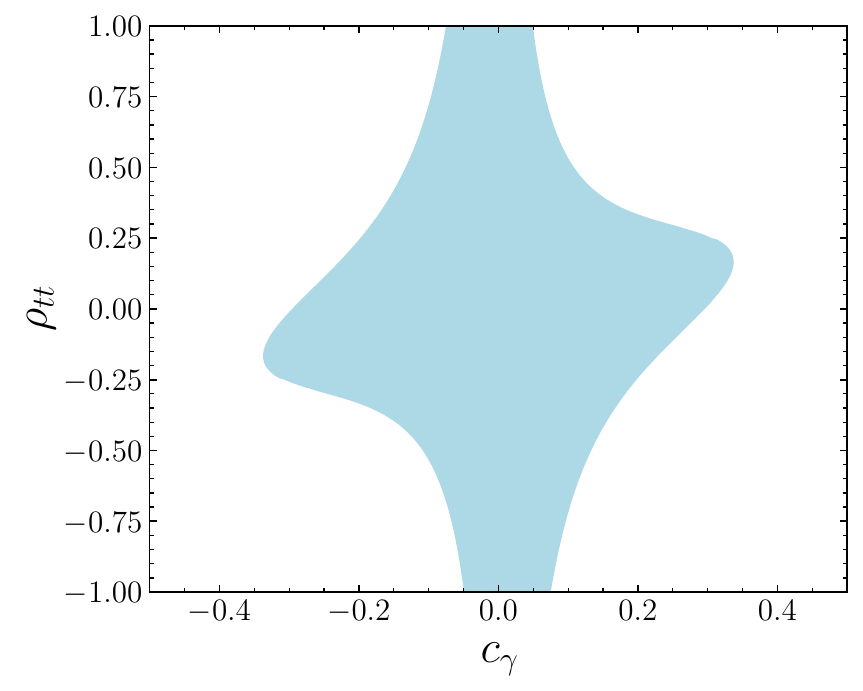}									
	\caption{The $2\sigma$ allowed parameter region by SM Higgs signal strength measurements.}	
	\label{fig:HS}
\end{figure}

At the experimental level, the most stringent constraints on the G2HDM arise from direct searches for additional Higgs bosons and precision measurements of the observed 125~GeV state at the LHC. Higgs signal strength measurements constrain the mixing angle~$c_\gamma$ between the two {\it CP}-even scalars $h$ and $H$, pushing the model close to the so-called alignment limit. Requiring the model to agree with the Higgs signal measurements constrains $c_\gamma$, $\rho_{tt}$ and $\eta_i$.\footnote{The $\eta_3$ and $\eta_7$ quartic couplings govern the interaction of $h$ with the charged Higgs $H^+$, thereby affecting the Higgs diphoton decay rate at one-loop level.} These parameters modify the SM-like Higgs couplings. Here, we identify the observed 125~GeV state with the light {\it CP}-even $h$. 
In Fig.~\ref{fig:HS}, we plot the $(c_\gamma$, $\rho_{tt})$ parameter space allowed by SM Higgs signal strength measurements, as obtained with \textsc{HiggsSignals} module of \textsc{HiggsTools}~\cite{Bahl:2022igd}. 
We find that values of $\rho_{tt}$ as large as 1 are consistent with Higgs data, provided that $c_\gamma$ remains sufficiently small. Conversely, $c_\gamma$ can reach its maximum ($\sim 0.3$) if $\rho_{tt}$ is small but non-vanishing, as long as it carries the same sign as $\rho_{tt}$.

\begin{figure}[t!]
	\centering
	\includegraphics[scale=0.5]{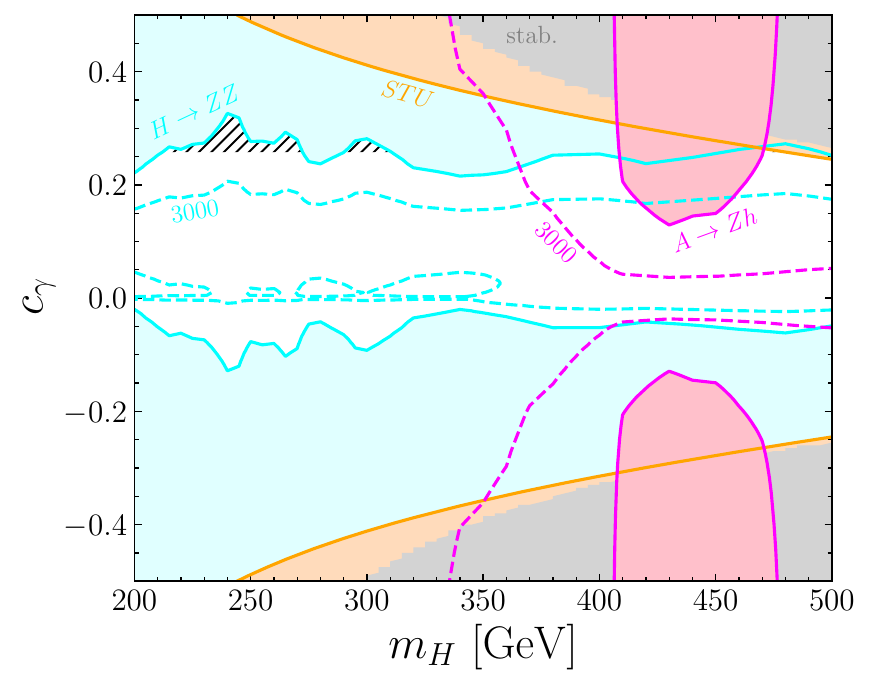}	
	\includegraphics[scale=0.5]{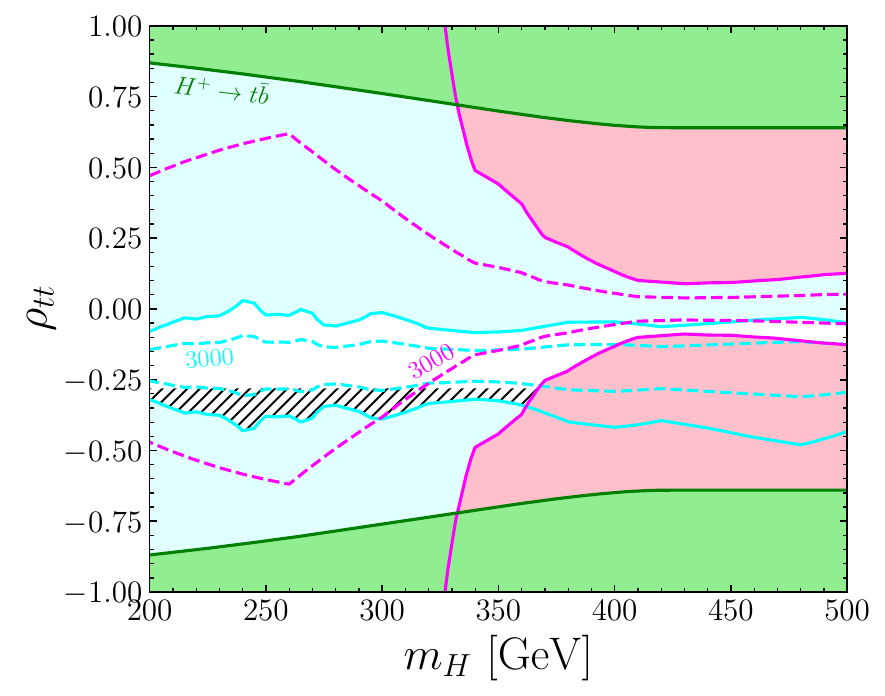}									
	\caption{Exclusion bounds in the $m_H$-$c_\gamma$ plane for $\rho_{tt} = -0.1$ (left) and in the $m_H$-$\rho_{tt}$ plane for $c_\gamma = 0.2$ (right) from vacuum stability requirement (light gray), electroweak precision constraints (orange), SM Higgs signal strength measurements (hatched), as well as $H \to ZZ$ (cyan), $A \to Zh$ (magenta), and $H^+ \to t\bar b$ (green) searches. The expected HL-LHC $H \to ZZ$ (dashed cyan) and $A \to Zh$ (dashed magenta) search limits are also shown.}	
	\label{fig:HB}
\end{figure}

Flavor observables, in particular $B_q$ ($q = d,s$) meson mixings and $b \to s\gamma$ decay, have already placed strong constraints on extra Yukawa couplings $\rho_{ij}$~\cite{Crivellin:2013wna}. Constraints on $\rho_{tt}$ are not particularly strong~\cite{Crivellin:2013wna,Altunkaynak:2015twa,Hou:2025tjp}. 
Here we assume $m_A = m_{H^+} = 500$~GeV and scan over $c_\gamma$, $\rho_{tt}$, and $m_H$ in the ranges: $c_\gamma \in [-0.5, 0.5]$, $\rho_{tt} \in [-1,1]$, $m_H \in [200, 500]~$GeV. The remaining parameters are chosen to be $\mu^2_{22}=m^2_H / 2$, $\eta_2 \simeq 2.52$ and $\eta_7 \simeq 0.17$ to satisfy theoretical restrictions discussed above.
In Fig.~\ref{fig:HB} we show the exclusion bounds on G2HDM parameter space.
In the left panel, the exclusion contours are shown in the $m_H$-$c_\gamma$ plane, assuming $\rho_{tt} = -0.1$.
In this scenario, the most sensitive limits come from searches for heavy resonances in the $H \to ZZ$~\cite{ATLAS:2020tlo} (cyan) and $A \to Zh$~\cite{ATLAS:2022enb} (magenta) decay channels. These exclusion limits are obtained using \textsc{HiggsBounds}\footnote{The ATLAS $A \to Zh$ search limit~\cite{ATLAS:2022enb} is not implemented in \textsc{HiggsBounds} and is thus applied by hand. The CMS limit~\cite{CMS:2024vxt}, which is relatively weaker, is not shown.} module of \textsc{HiggsTools}.
Hatched region is excluded by \textsc{HiggsSignals} (see Fig.~\ref{fig:HS}). Orange and light gray regions are excluded by electroweak precision constraints (through the oblique parameters $S$, $T$ and $U$) and by theoretical considerations of vacuum stability, respectively.    
Assuming $c_\gamma = 0.2$, we plot in the right panel of Fig.~\ref{fig:HB} the exclusion contours in the $m_H$-$\rho_{tt}$ plane. 
Analogous to the left panel, for small $\rho_{tt}$ coupling, the leading limits come from $H \to ZZ$~\cite{ATLAS:2020tlo} (cyan) and $A \to Zh$~\cite{ATLAS:2022enb} (magenta), 
while for large $\rho_{tt}$ coupling, $H^+ \to t\bar b$~\cite{ATLAS:2021upq} (green) and $A \to t\bar t$~\cite{CMS:2019pzc} (once the $t\bar t$ decay mode is open) are also sensitive. Limits from $H^+ \to W^+h$~\cite{ATLAS:2024rcu} searches are very weak.
The expected reach of the high-luminosity LHC (HL-LHC) is also shown in Fig.~\ref{fig:HB} based on a naive $\sqrt{\mathcal{L}}$ extrapolation.
We observe that the HL-LHC could probe $|c_\gamma|$ and $|\rho_{tt}|$ down to $\sim 0.1$. 

\section{Collider Analysis}\label{sect:collider}
\begin{figure}[t]
	\centering
	\includegraphics[scale=0.55]{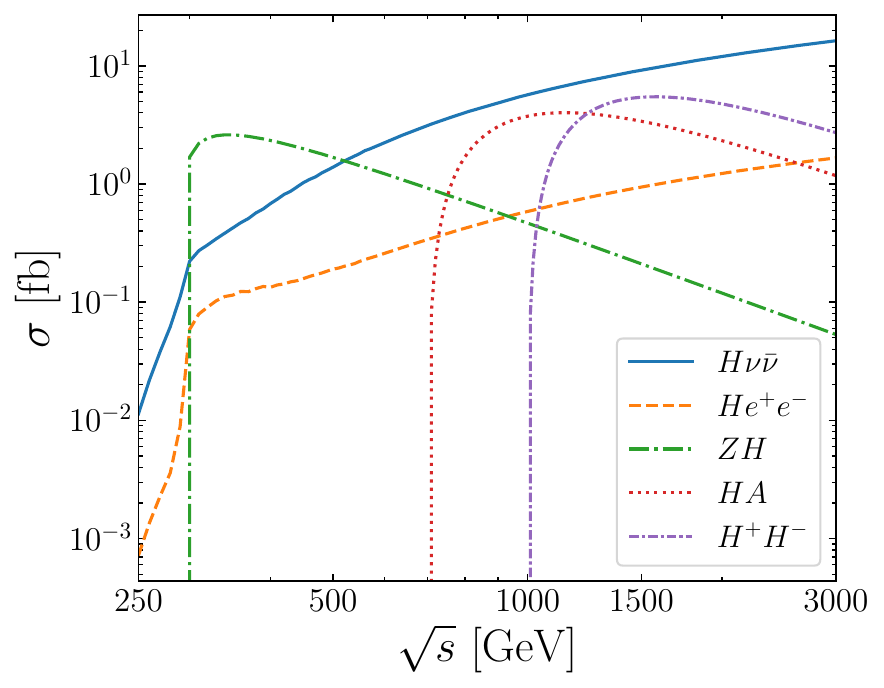}									
	\caption{Cross sections vs. $\sqrt{s}$ for $m_{H} = 200$~GeV, $m_A = m_{H^+} = 500$~GeV, and $c_\gamma = 0.2$. Cross sections are calculated at leading order using \textsc{MadGraph5\_aMC@NLO}~\cite{Alwall:2014hca}.}	
	\label{fig:CSs}
\end{figure}

We now discuss the discovery potential of the extra neutral Higgs boson $H$ at CLIC with 1.5 TeV collision energy and 4 ab$^{-1}$ luminosity.
The signal process is $e^+e^- \to H \nu \bar\nu \to W^+ W^- \nu \bar\nu \to \ell^+\ell^- \nu \bar \nu \nu \bar \nu$.
We consider $m_H = 200~(250)$ GeV, $m_A = m_{H^+} = 500$~GeV, $c_\gamma = 0.2$, and $\rho_{tt} = -0.1$ as a representative benchmark point, denoted BP1 (BP2), for which the predominant decay channel of $H$ is $H \to W^+ W^-$, followed by $H \to ZZ$.\footnote{The $H \to ZZ$ channel is not included in our analysis, but would lead to the same final state via $Z \to \ell^+ \ell^-$ and $Z \to \nu\bar\nu$.}
Note that only $WW$-fusion is considered for the $H$ production; production process through $ZZ$-fusion, i.e. $e^+ e^- \to H e^+e^-$, as well as $e^+e^- \to ZH$ and $e^+ e^- \to HA$ processes have relatively small production cross section at TeV energies (see Fig.~\ref{fig:CSs}).
The main SM backgrounds are electroweak boson pair
production processes $WW$ and $ZZ$, top quark pair production
$t\bar t$, and $\ell^+\ell^-$ production\footnote{For $\ell=e$, the process includes both $s$-channel and $t$-channel diagrams. For $\mu,\tau$ only $s$-channel $Z/\gamma^*$ exchange contributes.} (referred to as $Z/\gamma^*$), and $h\nu \bar\nu$ associated production. 
Other backgrounds such as $he^+e^-$ and $Zh$ are minor ($<1\%$ after selection cuts).
We recall that $h$ refers to the SM-like Higgs boson.
It should be noted that the incoming electron and positron beams are assumed to be unpolarized.

Signal and background events are sampled at leading order using  \textsc{MadGraph5\_aMC@NLO}~\cite{Alwall:2014hca}. 
These events are passed to \textsc{Pythia-8.2}~\cite{Sjostrand:2014zea} for parton showering and hadronization, and then to \textsc{Delphes-3.5.0} \cite{deFavereau:2013fsa}, for fast detector simulation, with the default \texttt{CLICdet\_Stage2} card~\cite{CLICdp:2018vnx}, where the jet clustering is performed using the Valencia algorithm~\cite{Boronat:2014hva,Boronat:2016tgd} via \textsc{FastJet}~\cite{Cacciari:2011ma}. 

\begin{figure*}[t]
	\centering
	\includegraphics[scale=0.5]{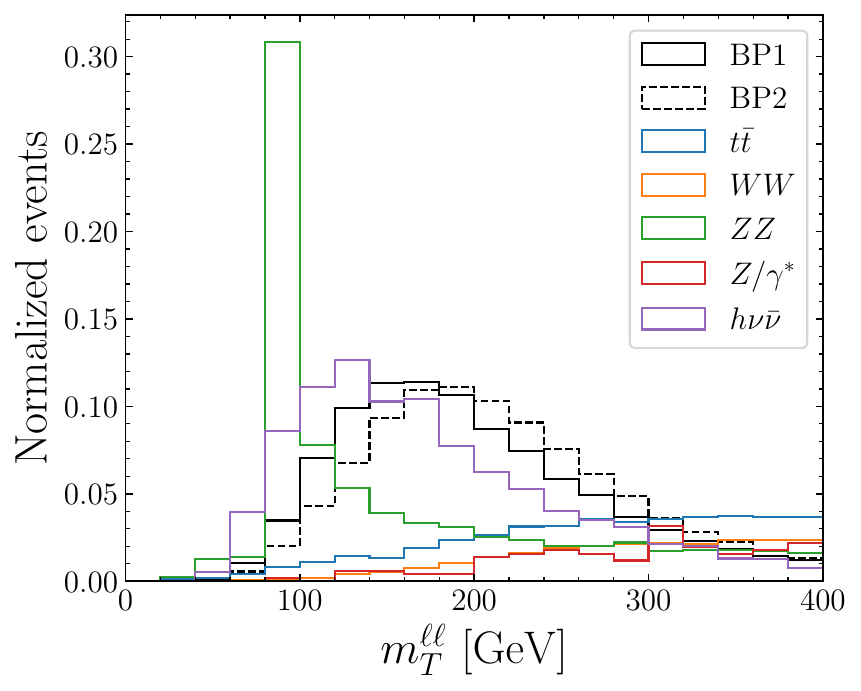}
	\includegraphics[scale=0.5]{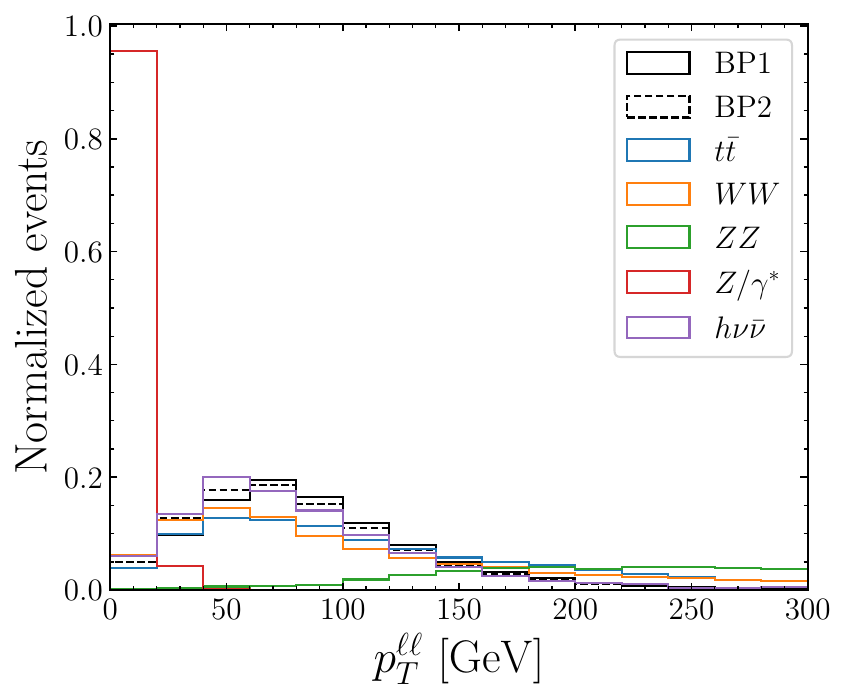}			
	\includegraphics[scale=0.5]{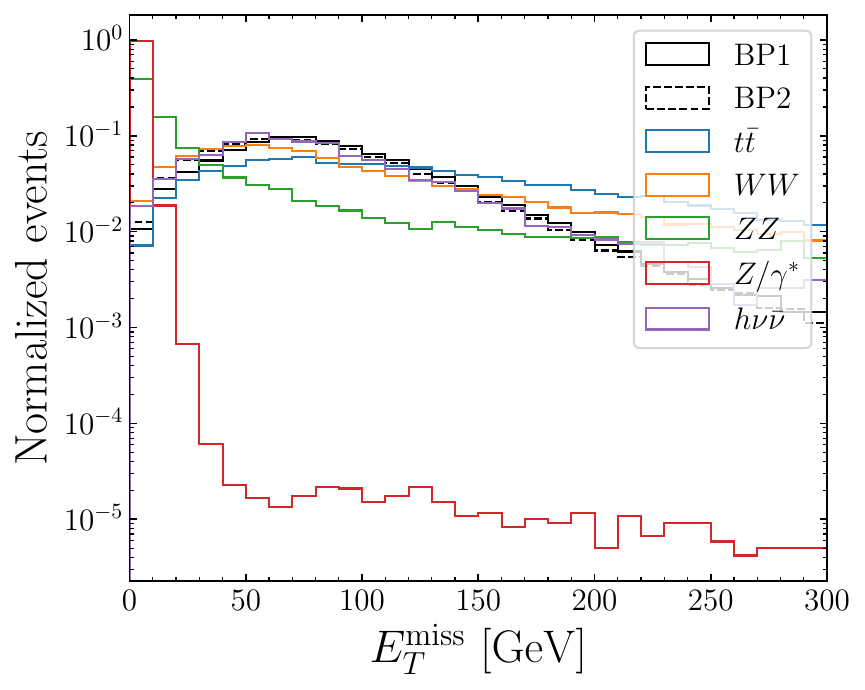}
	\caption{Normalized distributions for signal and background processes at 1.5~TeV CLIC.}	
	\label{fig:distr}
\end{figure*}

Events with exactly two opposite-sign (OS) leptons, with $p^{\ell}_T > 20$ and $|\eta_\ell|<2.5$, are selected. Events satisfying this requirement are dominated by the contribution from the $Z/\gamma^*$ process. To reduce this background, an additional requirement on the transverse mass $m^{\ell\ell}_T$ is applied.
The $m^{\ell\ell}_T$ variable is also effective in reducing contributions from the $t\bar t$, $WW$, and $ZZ$ processes.
The missing transverse energy, $E^{\rm miss}_T$, is used to further reduce the $ZZ$ background.
After applying the $m^{\ell\ell}_T$, and $E^{\rm miss}_T$ selections, the remaining events come mainly from $h\nu\bar\nu$, $t\bar t$, and $WW$ backgrounds. To suppress the $t\bar t$ contribution, events containing at least one jet candidate with $p_T > 25$~GeV are vetoed (referred to as ``jet veto'').
At this stage, the signal-to-background ratio is already significantly improved.
However, to further reduce the $Z/\gamma^*$ contribution, the dilepton transverse momentum $p^{\ell\ell}_T$ is required to be greater than 30~GeV.
These selection cuts are optimized to maximize the statistical significance of the signal.
The cross sections of simulated signal and background events remaining after each step of the event selection are summarized in Table~\ref{table:xsASC}.
Fig.~\ref{fig:distr} shows the $m^{\ell\ell}_T$, $p^{\ell\ell}_T$, and $E^{\rm miss}_T$ distributions before applying the selection cuts. 

\begin{table*}[b]
	\centering
	\setlength{\tabcolsep}{8pt}
	{\small \begin{tabular}{l c c c c c c c} 
			\hline\hline
			$\sqrt{s} = 1.5$~TeV & BP1 & BP2 & $t\bar t$ & $WW$ & $ZZ$ & $Z/\gamma^*$ & $h\nu \bar\nu$ \\\hline
            Exactly two OS leptons & 0.189 & 0.160 & 3.176 & 25.289 & 3.200 & 11013.7 & 1.089 \\		
            $120 < m^{\ell\ell}_T < 260$ GeV  & 0.124 & 0.104 & 0.469 & 1.362 & 0.669 & 0.304 & 0.618 \\	
            $E^{\rm{miss}}_T > 50$ GeV        & 0.104 & 0.074 & 0.373 & 0.295 & 0.055 & 0.120 & 0.569 \\			
            Jet veto                          & 0.104 & 0.074 & $<0.01$ & 0.284 & 0.023 & 0.120 & 0.555 \\ 
            $p^{\ell\ell}_T > 30$ GeV         & 0.104 & 0.074 & $<0.01$ & 0.263 & 0.023 & 0.101 & 0.553 \\ \hline\hline							
	\end{tabular}}
	\caption{Signal and background cross sections (in fb) at each event selection cut.} 
	\label{table:xsASC} 
\end{table*} 
\begin{figure*}[t]
	\centering
	\includegraphics[scale=0.5]{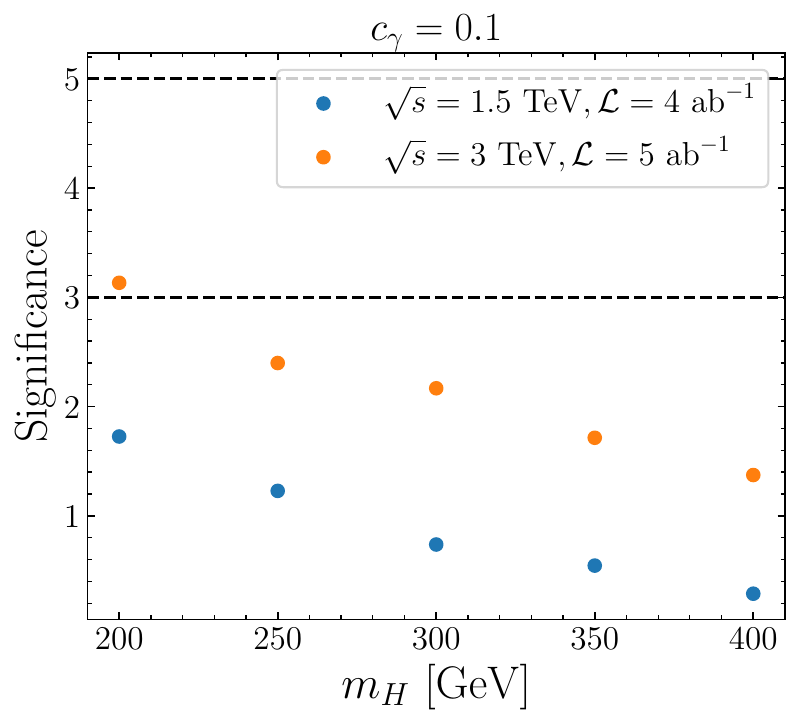}	
	\includegraphics[scale=0.5]{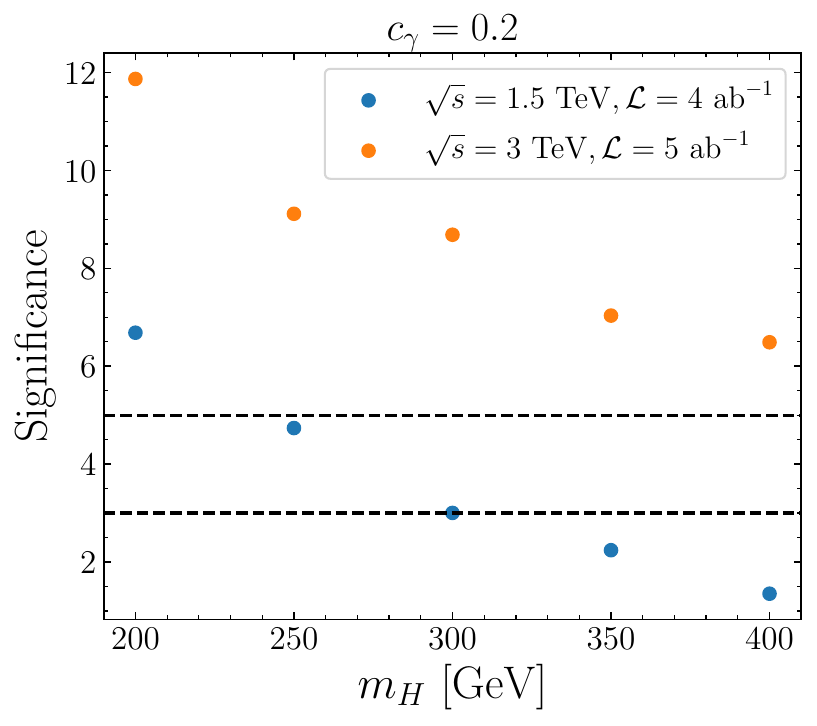}									
	\caption{Significance as a function of $m_H$ for both energy stages of CLIC.}	
	\label{fig:significance}
\end{figure*}

We now turn to estimate the signal significance from the cross sections summarized in Table~\ref{table:xsASC}. By using $\mathcal{Z} = \sqrt{2\left[(N_S+N_B)\ln(1+N_S/N_B)-N_S\right]}$~\cite{Cowan:2010js}, where $N_S$ ($N_B$) is the number of signal (background) events, and assuming a luminosity of 4~ab$^{-1}$, BP1 (BP2) yields a significance of $\sim 6.7\sigma$ (4.8$\sigma$).
These results illustrate that 1.5~TeV CLIC could be able to probe $m_H$ at ${\mathcal O}(200)$~GeV through $e^+e^- \to H\nu \bar\nu \to W^+W^-\nu\bar\nu$.
Applying the same selection cuts as in Table~\ref{table:xsASC}, except the dilepton transverse mass, which is required to satisfy $140 < m^{\ell\ell}_T < 340$~GeV, we find $\mathcal{Z} \sim 3.0\sigma$ for $m_H = 300$~GeV. 
Significance as a function of $m_H$ for both center-of-mass energies of 1.5 and 3~TeV is shown in Fig.~\ref{fig:significance}.

Before concluding, we note that the process $e^+ e^- \to H^+ H^-$ can also occur.
It was shown that $e^+ e^- \to H^+ H^- \to c\bar b \bar c b$ may allow the extraction of a charged Higgs signal at the ILC, with $m_H \sim m_A \sim m_{H^+} \sim 200$~GeV~\cite{Hou:2021qff}. The companion process $e^+ e^- \to HA \to t\bar c t\bar c, \bar t c\bar t c$ can give rise to same-sign top quark pair production~\cite{Hou:1995qh}.
The decays $H^+ \to c\bar b$ and $H/A \to t\bar c$ are both induced by the FCNH coupling $\rho_{tc}$, and the signals above may reveal the presence of G2HDM with flavor violation. 
Here, we have given the scenario of $c_\gamma \neq 0$ and weak $\rho_{tt}$ coupling, where the latter is still sufficient for EWBG~\cite{Fuyuto:2017ewj}. The FCNH $\rho_{tc}$ coupling is turned off in our study; however, turning it on may reduce the achievable significance illustrated in Fig.~\ref{fig:significance}. Assuming $\rho_{tc} = 0.1$, which evades bounds from $t \to ch$ search~\cite{ATLAS:2024mih,CMS:2024ubt} and HL-LHC search~\cite{Hou:2020chc}, would induce $H \to t\bar c$ and soften $H \to WW$ decay but still the latter is the predominant channel.
In our scenario, the presence of $A$ and $H^+$ can be revealed via $e^+e^- \to HA \to H ZH, HZh$ and $e^+e^- \to H^+H^- \to W^+H W^-H, W^+hW^-h$, respectively, but this would require further investigation, which is left for future work.

\section{Conclusion}\label{sect:concl}
We have discussed the possibility of probing an extra boson Higgs at future linear $e^+e^-$ colliders.
We first focused on constraining the $h$-$H$ mixing angle $c_\gamma$, and the extra top Yukawa coupling $\rho_{tt}$, the EWBG driver~\cite{Fuyuto:2017ewj}, using LHC data on the 125 GeV Higgs properties as well as searches for additional Higgs bosons. 
We show that HL-LHC would be able to probe $c_\gamma$ and $\rho_{tt}$ down to $\sim 0.1$ using $H \to ZZ$ and $A \to Zh$, with $h$ the SM-like Higgs boson. 
We then studied the process $e^+e^- \to H\nu \bar\nu \to W^+W^-\nu\bar\nu$ at 1.5 and 3~TeV CLIC, which allows a direct measurement of the Higgs mixing angle~$c_\gamma$. 
We demonstrated, through a detector-level signal-to-background analysis, that CLIC could probe an extra Higgs boson with masses in the range 200-400~GeV.

\section*{Acknowledgments}
This work is supported by the National Science and Technology Council of Taiwan under grant No.~114-2639-M-002-006-ASP, and NTU grants No.~114L86001 and No.~114L891801.
M.~K. would like to thank Leon M.G. de la Vega for many fruitful discussions and NCTS Physics Division for hospitality, where part of this work was carried out.

\bibliographystyle{JHEP}
\bibliography{mybib}

\providecommand{\href}[2]{#2}\begingroup\raggedright\begin{thebibliography}{10}

\bibitem{ATLAS:2012yve}
{\bf ATLAS} Collaboration, G.~Aad et~al., {\it {Observation of a new particle
  in the search for the Standard Model Higgs boson with the ATLAS detector at
  the LHC}},  {\em Phys. Lett. B} {\bf 716} (2012) 1--29,
  [\href{http://arxiv.org/abs/1207.7214}{{\tt arXiv:1207.7214}}].

\bibitem{CMS:2012qbp}
{\bf CMS} Collaboration, S.~Chatrchyan et~al., {\it {Observation of a New Boson
  at a Mass of 125 GeV with the CMS Experiment at the LHC}},  {\em Phys. Lett.
  B} {\bf 716} (2012) 30--61, [\href{http://arxiv.org/abs/1207.7235}{{\tt
  arXiv:1207.7235}}].

\bibitem{Fuyuto:2017ewj}
K.~Fuyuto, W.-S. Hou, and E.~Senaha, {\it {Electroweak baryogenesis driven by
  extra top Yukawa couplings}},  {\em Phys. Lett. B} {\bf 776} (2018) 402--406,
  [\href{http://arxiv.org/abs/1705.05034}{{\tt arXiv:1705.05034}}].

\bibitem{Hou:1991un}
W.-S. Hou, {\it {Tree level $t \to ch$ or $h \to t\bar{c}$ decays}},  {\em
  Phys. Lett. B} {\bf 296} (1992) 179--184.

\bibitem{Chen:2013qta}
K.-F. Chen, W.-S. Hou, C.~Kao, and M.~Kohda, {\it {When the Higgs meets the
  Top: Search for $t \to ch^0$ at the LHC}},  {\em Phys. Lett. B} {\bf 725}
  (2013) 378--381, [\href{http://arxiv.org/abs/1304.8037}{{\tt
  arXiv:1304.8037}}].

\bibitem{ATLAS:2024mih}
{\bf ATLAS} Collaboration, G.~Aad et~al., {\it {Search for flavour-changing
  neutral-current couplings between the top quark and the Higgs boson in
  multi-lepton final states in 13~TeV pp collisions with the ATLAS detector}},
  {\em Eur. Phys. J. C} {\bf 84} (2024), no.~7 757,
  [\href{http://arxiv.org/abs/2404.02123}{{\tt arXiv:2404.02123}}].

\bibitem{CMS:2024ubt}
{\bf CMS} Collaboration, A.~Hayrapetyan et~al., {\it {Search for
  flavor-changing neutral current interactions of the top quark mediated by a
  Higgs boson in proton-proton collisions at 13 TeV}},  {\em Phys. Rev. D} {\bf
  112} (2025), no.~3 032008, [\href{http://arxiv.org/abs/2407.15172}{{\tt
  arXiv:2407.15172}}].

\bibitem{Hou:2024bzh}
W.-S. Hou and M.~Krab, {\it {Reconstructing the general 2HDM charged Higgs
  boson at the LHC}},  {\em Phys. Rev. D} {\bf 110} (2024), no.~1 L011702,
  [\href{http://arxiv.org/abs/2405.19190}{{\tt arXiv:2405.19190}}].

\bibitem{Krab:2025zuy}
M.~Krab, {\it {Searching for charged Higgs bosons with flavor-changing
  couplings at the LHC}},  in {\em {13th Large Hadron Collider Physics
  Conference}}, 8, 2025.
\newblock \href{http://arxiv.org/abs/2508.00764}{{\tt arXiv:2508.00764}}.

\bibitem{Linssen:2012hp}
{\it {Physics and Detectors at CLIC: CLIC Conceptual Design Report}},
  \href{http://arxiv.org/abs/1202.5940}{{\tt arXiv:1202.5940}}.

\bibitem{Aicheler:2018arh}
{\bf CLIC accelerator} Collaboration, {\it {The Compact Linear Collider (CLIC)
  - Project Implementation Plan}},  \href{http://arxiv.org/abs/1903.08655}{{\tt
  arXiv:1903.08655}}.

\bibitem{ILCInternationalDevelopmentTeam:2022izu}
{\bf ILC International Development Team} Collaboration, A.~Aryshev et~al., {\it
  {The International Linear Collider: Report to Snowmass 2021}},
  \href{http://arxiv.org/abs/2203.07622}{{\tt arXiv:2203.07622}}.

\bibitem{FCC:2025lpp}
{\bf FCC} Collaboration, M.~Benedikt et~al., {\it {Future Circular Collider
  Feasibility Study Report: Volume 1, Physics, Experiments, Detectors}},
  \href{http://arxiv.org/abs/2505.00272}{{\tt arXiv:2505.00272}}.

\bibitem{CEPCPhysicsStudyGroup:2022uwl}
{\bf CEPC Physics Study Group} Collaboration, H.~Cheng et~al., {\it {The
  Physics potential of the CEPC. Prepared for the US Snowmass Community
  Planning Exercise (Snowmass 2021)}},  in {\em {Snowmass 2021}}, 5, 2022.
\newblock \href{http://arxiv.org/abs/2205.08553}{{\tt arXiv:2205.08553}}.

\bibitem{Abramowicz:2016zbo}
H.~Abramowicz et~al., {\it {Higgs physics at the CLIC
  electron{\textendash}positron linear collider}},  {\em Eur. Phys. J. C} {\bf
  77} (2017), no.~7 475, [\href{http://arxiv.org/abs/1608.07538}{{\tt
  arXiv:1608.07538}}].

\bibitem{CLICdp:2018esa}
{\bf CLICdp} Collaboration, H.~Abramowicz et~al., {\it {Top-Quark Physics at
  the CLIC Electron-Positron Linear Collider}},  {\em JHEP} {\bf 11} (2019)
  003, [\href{http://arxiv.org/abs/1807.02441}{{\tt arXiv:1807.02441}}].

\bibitem{CLIC:2018fvx}
{\bf CLIC} Collaboration, J.~de~Blas et~al., {\it {The CLIC Potential for New
  Physics}},  {\em CERN Yellow Rep. Monogr.} {\bf 3} (2018) 1--282,
  [\href{http://arxiv.org/abs/1812.02093}{{\tt arXiv:1812.02093}}].

\bibitem{Cheung:2023qnj}
K.~Cheung, Y.-n. Mao, S.~Moretti, and R.~Zhang, {\it {Testing CP-violation in a
  heavy Higgs sector at CLIC}},  {\em Eur. Phys. J. C} {\bf 85} (2025), no.~6
  700, [\href{http://arxiv.org/abs/2304.04390}{{\tt arXiv:2304.04390}}].

\bibitem{Hashemi:2023tej}
M.~Hashemi and M.~Molanaei, {\it {Heavy neutral 2HDM Higgs boson pair
  production at CLIC energies}},  {\em Phys. Rev. D} {\bf 108} (2023), no.~3
  035012, [\href{http://arxiv.org/abs/2306.16116}{{\tt arXiv:2306.16116}}].

\bibitem{Lee:2025hgb}
S.~Lee, D.~Kim, J.-H. Cho, J.~Kim, and J.~Song, {\it {Multi-step Strong
  First-Order Electroweak Phase Transitions in the Inverted Type-I 2HDM:
  Parameter Space, Gravitational Waves, and Collider Phenomenology}},
  \href{http://arxiv.org/abs/2506.03260}{{\tt arXiv:2506.03260}}.

\bibitem{Davidson:2005cw}
S.~Davidson and H.~E. Haber, {\it {Basis-independent methods for the
  two-Higgs-doublet model}},  {\em Phys. Rev. D} {\bf 72} (2005) 035004,
  [\href{http://arxiv.org/abs/hep-ph/0504050}{{\tt hep-ph/0504050}}]. [Erratum:
  Phys.Rev.D 72, 099902 (2005)].

\bibitem{Hou:2017hiw}
W.-S. Hou and M.~Kikuchi, {\it {Approximate Alignment in Two Higgs Doublet
  Model with Extra Yukawa Couplings}},  {\em EPL} {\bf 123} (2018), no.~1
  11001, [\href{http://arxiv.org/abs/1706.07694}{{\tt arXiv:1706.07694}}].

\bibitem{Bahl:2022igd}
H.~Bahl, T.~Biek{\"o}tter, S.~Heinemeyer, C.~Li, S.~Paasch, G.~Weiglein, and
  J.~Wittbrodt, {\it {HiggsTools: BSM scalar phenomenology with new versions of
  HiggsBounds and HiggsSignals}},  {\em Comput. Phys. Commun.} {\bf 291} (2023)
  108803, [\href{http://arxiv.org/abs/2210.09332}{{\tt arXiv:2210.09332}}].

\bibitem{Crivellin:2013wna}
A.~Crivellin, A.~Kokulu, and C.~Greub, {\it {Flavor-phenomenology of
  two-Higgs-doublet models with generic Yukawa structure}},  {\em Phys. Rev. D}
  {\bf 87} (2013), no.~9 094031, [\href{http://arxiv.org/abs/1303.5877}{{\tt
  arXiv:1303.5877}}].

\bibitem{Altunkaynak:2015twa}
B.~Altunkaynak, W.-S. Hou, C.~Kao, M.~Kohda, and B.~McCoy, {\it {Flavor
  Changing Heavy Higgs Interactions at the LHC}},  {\em Phys. Lett. B} {\bf
  751} (2015) 135--142, [\href{http://arxiv.org/abs/1506.00651}{{\tt
  arXiv:1506.00651}}].

\bibitem{Hou:2025tjp}
W.-S. Hou and M.~Krab, {\it {Probing the general 2HDM with flavor violation
  through A{\textrightarrow}ZH}},  {\em Phys. Rev. D} {\bf 111} (2025), no.~11
  115036, [\href{http://arxiv.org/abs/2503.23133}{{\tt arXiv:2503.23133}}].

\bibitem{ATLAS:2020tlo}
{\bf ATLAS} Collaboration, G.~Aad et~al., {\it {Search for heavy resonances
  decaying into a pair of Z bosons in the $\ell ^+\ell ^-\ell '^+\ell '^-$ and
  $\ell ^+\ell ^-\nu {{\bar{\nu }}}$ final states using 139 $\mathrm {fb}^{-1}$
  of proton{\textendash}proton collisions at $\sqrt{s} = 13\,$TeV with the
  ATLAS detector}},  {\em Eur. Phys. J. C} {\bf 81} (2021), no.~4 332,
  [\href{http://arxiv.org/abs/2009.14791}{{\tt arXiv:2009.14791}}].

\bibitem{ATLAS:2022enb}
{\bf ATLAS} Collaboration, G.~Aad et~al., {\it {Search for heavy resonances
  decaying into a $Z$ or $W$ boson and a Higgs boson in final states with
  leptons and $b$-jets in $139~$fb$^{-1}$ of $pp$ collisions at
  $\sqrt{s}=13~$TeV with the ATLAS detector}},  {\em JHEP} {\bf 06} (2023) 016,
  [\href{http://arxiv.org/abs/2207.00230}{{\tt arXiv:2207.00230}}].

\bibitem{CMS:2024vxt}
{\bf CMS} Collaboration, {\it {Search for a heavy CP-odd Higgs boson decaying
  into a 125 GeV Higgs boson and a Z boson in final states with two tau and two
  light leptons at sqrts = 13 TeV}}, .

\bibitem{ATLAS:2021upq}
{\bf ATLAS} Collaboration, G.~Aad et~al., {\it {Search for charged Higgs bosons
  decaying into a top quark and a bottom quark at $ \sqrt{\mathrm{s}} $ = 13
  TeV with the ATLAS detector}},  {\em JHEP} {\bf 06} (2021) 145,
  [\href{http://arxiv.org/abs/2102.10076}{{\tt arXiv:2102.10076}}].

\bibitem{CMS:2019pzc}
{\bf CMS} Collaboration, A.~M. Sirunyan et~al., {\it {Search for heavy Higgs
  bosons decaying to a top quark pair in proton-proton collisions at $\sqrt{s}
  =$ 13 TeV}},  {\em JHEP} {\bf 04} (2020) 171,
  [\href{http://arxiv.org/abs/1908.01115}{{\tt arXiv:1908.01115}}]. [Erratum:
  JHEP 03, 187 (2022)].

\bibitem{ATLAS:2024rcu}
{\bf ATLAS} Collaboration, G.~Aad et~al., {\it {Search for a heavy charged
  Higgs boson decaying into a W boson and a Higgs boson in final states with
  leptons and b-jets in $ \sqrt{s} $ = 13 TeV pp collisions with the ATLAS
  detector}},  {\em JHEP} {\bf 02} (2025) 143,
  [\href{http://arxiv.org/abs/2411.03969}{{\tt arXiv:2411.03969}}].

\bibitem{Alwall:2014hca}
J.~Alwall, R.~Frederix, S.~Frixione, V.~Hirschi, F.~Maltoni, O.~Mattelaer,
  H.~S. Shao, T.~Stelzer, P.~Torrielli, and M.~Zaro, {\it {The automated
  computation of tree-level and next-to-leading order differential cross
  sections, and their matching to parton shower simulations}},  {\em JHEP} {\bf
  07} (2014) 079, [\href{http://arxiv.org/abs/1405.0301}{{\tt
  arXiv:1405.0301}}].

\bibitem{Sjostrand:2014zea}
T.~Sj{\"o}strand, S.~Ask, J.~R. Christiansen, R.~Corke, N.~Desai, P.~Ilten,
  S.~Mrenna, S.~Prestel, C.~O. Rasmussen, and P.~Z. Skands, {\it {An
  introduction to PYTHIA 8.2}},  {\em Comput. Phys. Commun.} {\bf 191} (2015)
  159--177, [\href{http://arxiv.org/abs/1410.3012}{{\tt arXiv:1410.3012}}].

\bibitem{deFavereau:2013fsa}
{\bf DELPHES 3} Collaboration, J.~de~Favereau, C.~Delaere, P.~Demin,
  A.~Giammanco, V.~Lema{\^\i}tre, A.~Mertens, and M.~Selvaggi, {\it {DELPHES 3,
  A modular framework for fast simulation of a generic collider experiment}},
  {\em JHEP} {\bf 02} (2014) 057, [\href{http://arxiv.org/abs/1307.6346}{{\tt
  arXiv:1307.6346}}].

\bibitem{CLICdp:2018vnx}
{\bf CLICdp} Collaboration, D.~Arominski et~al., {\it {A detector for CLIC:
  main parameters and performance}},
  \href{http://arxiv.org/abs/1812.07337}{{\tt arXiv:1812.07337}}.

\bibitem{Boronat:2014hva}
M.~Boronat, J.~Fuster, I.~Garcia, E.~Ros, and M.~Vos, {\it {A robust jet
  reconstruction algorithm for high-energy lepton colliders}},  {\em Phys.
  Lett. B} {\bf 750} (2015) 95--99, [\href{http://arxiv.org/abs/1404.4294}{{\tt
  arXiv:1404.4294}}].

\bibitem{Boronat:2016tgd}
M.~Boronat, J.~Fuster, I.~Garcia, P.~Roloff, R.~Simoniello, and M.~Vos, {\it
  {Jet reconstruction at high-energy electron{\textendash}positron colliders}},
   {\em Eur. Phys. J. C} {\bf 78} (2018), no.~2 144,
  [\href{http://arxiv.org/abs/1607.05039}{{\tt arXiv:1607.05039}}].

\bibitem{Cacciari:2011ma}
M.~Cacciari, G.~P. Salam, and G.~Soyez, {\it {FastJet User Manual}},  {\em Eur.
  Phys. J. C} {\bf 72} (2012) 1896, [\href{http://arxiv.org/abs/1111.6097}{{\tt
  arXiv:1111.6097}}].

\bibitem{Cowan:2010js}
G.~Cowan, K.~Cranmer, E.~Gross, and O.~Vitells, {\it {Asymptotic formulae for
  likelihood-based tests of new physics}},  {\em Eur. Phys. J. C} {\bf 71}
  (2011) 1554, [\href{http://arxiv.org/abs/1007.1727}{{\tt arXiv:1007.1727}}].
  [Erratum: Eur.Phys.J.C 73, 2501 (2013)].

\bibitem{Hou:2021qff}
W.-S. Hou, R.~Jain, and T.~Modak, {\it {Searching for Charged Higgs Bosons via
  $e^+ e^- \to H^+ H^- \to c\bar b \bar c b$ at Linear Colliders}},  {\em JHEP}
  {\bf 07} (2022) 137, [\href{http://arxiv.org/abs/2111.06523}{{\tt
  arXiv:2111.06523}}].

\bibitem{Hou:1995qh}
W.-S. Hou and G.-L. Lin, {\it {Like sign top quark pair production at linear
  colliders}},  {\em Phys. Lett. B} {\bf 379} (1996) 261--266,
  [\href{http://arxiv.org/abs/hep-ph/9510359}{{\tt hep-ph/9510359}}].

\bibitem{Hou:2020chc}
W.-S. Hou and T.~Modak, {\it {Probing Top Changing Neutral Higgs Couplings at
  Colliders}},  {\em Mod. Phys. Lett. A} {\bf 36} (2021), no.~07 2130006,
  [\href{http://arxiv.org/abs/2012.05735}{{\tt arXiv:2012.05735}}].

\end{thebibliography}\endgroup
	
\end{document}